\documentstyle[twocolumn,epsfig,prl,aps,amsmath,amssymb]{revtex}

\begin{document}
\twocolumn[\hsize\textwidth\columnwidth\hsize\csname @twocolumnfalse\endcsname

\title{Dislocation lines as the precursor of the melting of crystalline solids
observed in Monte Carlo simulations}
\author{L. G\'omez$^{1,2,I}$, A. Dobry$^{1,2,II}$, Ch. Geuting$^2$, 
H. T. Diep$^{2,III}$ and L. Burakovsky$^{3,IV}$}
\address{$^1$ Facultad de Ciencias Exactas Ingenieria y Agrimensura, \\
and Instituto de F\'{\i}sica Rosario, Avenida Pellegrini 250,\\
2000 Rosario, Argentina\\
$^2$Laboratoire de Physique Th\'eorique et Mod\'elisation,\\
CNRS-Universit\'e de Cergy-Pontoise,\\
5, mail Gay-Lussac, Neuville sur Oise, 95031 Cergy-Pontoise Cedex, France\\
$^3$ Theoretical Division, LANL, Los Alamos, NM 87545, USA}
\date{\today}
\maketitle
\begin{abstract}
The microscopic mechanism of the melting of a crystal is analyzed by 
the constant pressure Monte Carlo simulation of a Lennard-Jones fcc system.
Beyond a temperature of the order of 0.8 of the melting temperature, we 
found that the relevant excitations are lines of defects. Each of these 
lines has the structure of a random walk of various lengths on an fcc defect 
lattice. We identify these lines with the dislocation ones proposed in recent 
phenomenological theories of melting. Near melting we find the appearance of 
long lines that cross the whole system. We suggest that these long lines are 
the precursor of the melting process.  
\end{abstract}
\vskip2pc] 
Melting is one of the rare phase transitions that can be observed in real 
life, outside of laboratories. Being a common-life process, the melting 
mechanism has been of interest for centuries. However there is yet no 
complete understanding of the atomistic dynamics involved in the melting 
transition. This is due to several difficulties found both in the experimental 
and theoretical studies of this problem. Let us discuss some of these 
difficulties.

Upon a phase transition long-range order found in the low temperature phase 
(LTP) disappears at the transition temperature. In the simplest cases, such as
a structural phase transition, order is associated with a geometrical quantity
which distinguishes LTP from the high temperature phase (HTP). The dynamical 
collective structural deformation, namely phonons, converting LTP into HTP is 
already present in the higher-symmetry phase. It is therefore natural to 
assume that the softening of this phonon excitation is the essential mechanism 
of the phase transition capturing the most important dynamics of the particles 
near the transition point.

However, at the melting temperature $T_m$, 
{\it both} translational and rotational symmetries of  
a crystal are destroyed, and it is much more complicated to construct simple 
models including the relevant excitations on both sides of the transition 
temperature. Hence, one-phase models have been developed. Starting in 
the solid phase, the question is what kind of excitation could destroy 
crystalline order. It is easy to see \cite{Vesh} that phonons alone cannot 
convert a solid into a liquid, some kind of crystalline defects should be 
invoked. Kosterlitz and Thouless \cite{KT} proposed a fundamental theory of 
the thermal breakdown of long-range order in two dimensions (2D) by 
topological defects, and related it to transitions in 2D crystals, 
superfluids and magnets, the relevant topological defects in the case of 
melting being crystalline dislocations (which are point defects in 2D). 
Their theory was greatly extended and detailed by Halperin and Nelson 
\cite{HN} and Yound \cite{Young} who predicted
that the complete transition from solid to liquid takes place in two steps: 
the dissociation of dislocation pairs drives a crystal into a liquid-crystal 
phase that retains finite-range orientational order, then a second transition 
at higher temperature completes the conversion to an isotropic liquid. This 
complete theory gave detailed predictions of the behavior of the specific 
heat, the structure factor, and elastic constants, that have been confirmed 
in numerous experiments and computer simulations. 

In three dimensions, the most reliable theories suggest that the defects that 
break crystalline order are dislocation lines, and that these lines 
proliferate at the melting temperature \cite{Kleinert,Bura}. In addition to 
big theoretical problems involved in the development of a defects-mediated 
melting model, experimental evidence showing thermal excitations of such 
dislocation lines is scarce if not non-existing \cite{experimental}. In fact, 
most of the thermodynamic properties of a crystal could be associated with 
phonons and their interactions up to temperatures very near the melting point 
\cite{Katsnelson}; the premelting temperature zone where the thermal 
excitations of defect lines should appear seems to be very narrow and 
difficult to access experimentally. 

Numerical simulations of model systems offer an alternative tool to
investigate this problem. In fact, from the beginning of the computational
era, the problem of melting has been studied by Monte Carlo (MC) and Molecular
Dynamics techniques. The equation of state including the melting points of
different materials has been obtained by these techniques, but the mechanism
underlying the solid-liquid transition is yet unclear. The essential problem
has been to separate the non-vibrational dynamics and to identify a premelting
zone where the excitations of some kind of defects prelude the breakdown of
crystalline order.

Here we return to this long-standing problem and try to shed new light on it. 
By constant-pressure MC simulation carried out for a Lennard-Jones (LJ) fcc 
crystal we follow the evolution of thermally activated defects. We show that 
local defects group in clusters which we identify as dislocation lines. These 
lines are rare at low temperature and do not contribute to the thermodynamic
properties of the system. However, as the temperature is increased, we succeed
in identifying a crossover temperature (called the premelting temperature, 
$T_{pm}$), of the order of $\sim 0.8\;T_{m}$. For $T>T_{pm}$ defect lines of 
all lengths are thermally activated and become the relevant excitation of 
the system. Moreover, very near the melting point we detect very long defect 
lines with the maximum avaible length of our simulated system. We conclude 
that these system-size-long defect lines are responsible for the melting 
transition.

Our simulations have been done on a cubic box of differents numbers of 
particles (between 864 up to 6912) interacting via a LJ  potential written as 
$V(r)=4\epsilon [(\frac{\sigma }{r})^{12}-(\frac{\sigma }{r})^{6}]$. 
Here $\epsilon $ is taken as the energy unit,
and $\sigma $ is fixed in such a way that the fcc lattice constant is equal
to 1 when only nearest-neighbor (NN) interaction is taken into account. We use 
periodic boundary conditions (PBC) numerically implemented by the minimal 
image convention method. It has been previously shown that in real systems 
melting starts at the surface of the sample \cite{Chokappa}. As our simulated 
sample does not contain a free surface the transition temperature may not 
correspond to the thermodynamic melting point. Nevertheless, we use 
PBC because this leads to an intrinsic bulk mechanism for melting, and 
the results are not strongly affected by the finite-size effects.

\begin{figure}[htb]
\vspace{1cm}
\epsfig{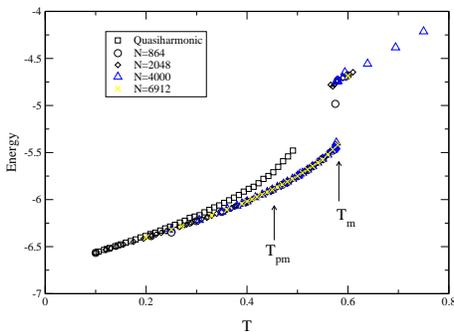}
\caption{Internal energy vs.\ $T$ for the system of $N=864,2048,4000$
and $6912$. $T_m$ and
$T_{pm}$ are the melting and pre-melting temperatures. See the main text
for the explanation.}
\label{evst}
\end{figure}

Let us start with the analysis of the internal energy of the system. 
In Fig.\ \ref{evst} we show this quantity as a function of temperature. 
A jump seen at $T_m=0.56\pm 0.01$ is associated with melting. For comparison 
we show our result of a quasiharmonic (QH) calculation on the same system. 
Both MC and QH calculations are done at zero pressure with a cutoff for the 
interaction taken at the next-nearest neighbors (NNN).  
We can see that the system behaves as harmonic only at low temperatures. The
anharmonicity of the interaction potential produces only a small dilation of
the system, but the essential dynamics of the particles is the harmonic
oscillations around the lattice sites. This is true almost up to 
temperatures of $\sim 0.4\;T_m$. For higher temperatures, phonons start to 
interact between themselves, and the anharmonicity becomes relevant. 
A refinement of the perturbative calculation based on phonons has been 
recently developed in Ref.\ \cite{Cowley}. 
Applying it to a LJ system with only NN interaction the authors of Ref.\ 
\onlinecite{Cowley} have shown that this theory can 
account for the thermodynamics properties of the system up to $\sim 0.8\;
T_{m}$. This fact implies that thermally excited defects, if  
present, do not contribute to the equilibrium properties up to this 
temperature (which we will call $T_{pm}$ in the following). $T_{pm}$ could be 
identified in Fig.\ \ref{evst} as the temperature where the energy starts to 
depart from the linear behavior. From $T_{pm}$ some other excitations are 
created, and we will study in detail this process in what follows. 

\begin{figure}[htp]
\vspace{1cm}
\epsfig{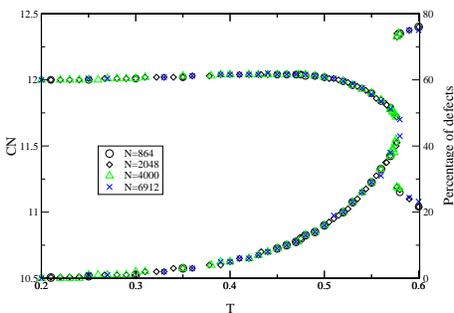}
\caption{The mean coordination number (left scale) and the
percentage of defects (right scale) as a function of T for
different cluster sizes.}
\label{CNvsT}
\end{figure}

We define a defect as a particle with a coordination number (CN) different 
from $12$ (the number of NN in an ideal fcc lattice). CN is obtained by 
counting the number of particles around a given one up to a given cutoff 
called $C_{NN}$. This cutoff is chosen as the value where the radial
distribution function 
has its first minimum. In Fig.\ \ref{CNvsT} 
we show the evolution with temperature of the average CN of the whole system. 
In agreement with our previous discussion, a considerable number of defective 
atoms appear at $T_{pm}$ where CN starts to decrease from 12. In this figure 
we also show the important increase of the percentage of defects with respect 
to the total number of atoms at $T_{pm}$. As no qualitative changes are 
observed for system sizes greather than $N=864$, in the following we will 
show results for the $N=2048$ system alone. 

In the present work we are interested in the correlation between defects, 
and in the study on how these defects group in clusters when the system 
gets closer to the melting point. We have developed the following algorithm 
to separate the defects in clusters and analyze their structure. We start 
with a given defect (let us call it $d_1$) and search new defects up to 
the cutoff distance $C_{NN}$. For each of these defects we repeat the same 
procedure. We iterate this process up to the completion of a cluster of 
connected defects. We label this cluster by the name of the first particle, 
$d_1.$ Then we take a new defect ($d_2$) disconnected 
from all of the previous ones and follow the same procedure. At the end 
of this process we obtain $N_{cl}$ disconnected clusters of defects.

By carefully analysing the distribution of distances {\it within} each 
cluster, we have determined its the internal geometry. When the clusters are 
relativelly dilute, our results indicate that the distance between the nearest 
and next nearest neighbors in a cluster are $\frac{\sqrt{2}}{2}$ and 
$\frac{\sqrt{3}}{2}$, respectivelly. As $\frac{\sqrt{3}}{2}$ is the distance 
between the vertex and the center of a unit cubic cell, we interpret 
the internal structure of the clusters as the system of (the parallel lines 
of) vacancies and interstitials. In every fcc cell inside the cluster one 
particle shifts from a vertex to the center leaving its perfect-lattice 
site. The new neighbors of this defective particle, at the centers of 
the neighboring faces of the cell, relax to stabilize this configuration, 
so that their separation corresponds to the equilibrium distance of 
the interaction potential. The resulting configuration is nothing but 
a vacancy-interstitial pair, which is the building block of the clusters 
of defects we observe. 

At low temperature these clusters are mainly isolated pairs of defects
of the type discussed above. When the temperature increases the density 
of defects increases as well. The creation of a pair of a vacancy and 
an intersticial  in the neighboring cells will be energetically more 
advantageous than their creation at a longer distance. This simple effect 
could be the origin of the excitation of chains of defects instead of isolated 
ones. These chains are the realization of the dislocation lines proposed in
phenomenological theories of melting \cite{Kleinert,Bura}, as we discuss in 
more detail below. It is therefore very important to check if these strings 
of defects appear in our simulation, and how they evolve near melting.

\begin{figure}[htb]
\epsfig{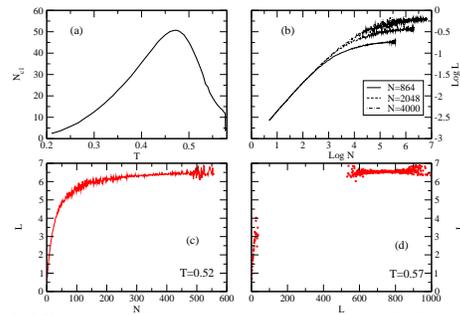}
\caption{(a) The average number of clusters as a function of $T$ for 
the system of 2048 particles. 
(b) The logarithm of the length $L$ of a cluster as a function of
the logarithm of the number $N$ of constuitutive particles of this cluster
at $T=0.52$ for three different system sizes. (c) $L$ vs.\ $N$ at $%
T=0.52$ for the system with $8\times 8\times 8$ cells. (d) The same for
$T=0.57$.}
\label{NvsL}
\end{figure}

We are interested in the behavior near melting of the 
mean total number of clusters (independent of their lengths) 
$N_{cl}$, and also in the dependence of the length $L$
(defined as the maximum distance between two particles within a cluster)
of a cluster on the number of particles $N$ within this cluster. 
In Fig.\ \ref{NvsL}a we show $N_{cl}$ as a function of $T$. The decrease of 
this quantity at $T_{pm}$ should be compared with the increase of the number 
of defective particles showed in Fig.\ \ref{CNvsT}. These facts indicate that 
the clusters are becoming bigger and bigger as melting is approached. 
In Fig.\ \ref{NvsL}b we show a plot of $\log L$ vs.\ $\log N$ for different 
system sizes at $T=0.52$, a temperature in the premelting zone. All these 
curves start with the same linear behavior. This indicates a power-law 
relationship between $L$ and $N$, of the form $L \sim N^\nu$ with 
$\nu\sim 0.62$. This value is in agreement with the law expected for 
a self-avoiding random walk (SAW) of a particle on the nodes of a 3D lattice 
\cite{geuting}. Therefore, what we see here is the development of strings of 
defects located at the sites of an fcc lattice. 
As far as we know, this is the first evidence that strings of the thermally 
excited defects are seen in a numerical simulation of a crystalline system.

The curves of Fig.\ \ref{NvsL}b saturate for large $N$. 
They tend to a value of $L$ (identified as $L_{s}$) which increases, as 
expected, when the size of the system increases. In the part (c) of the figure
we show the behavior of $L$ vs.\ $N$ for the system of $8\times 8\times 8$ 
unit cells. In this case $L_{s}$ is equal to $4\sqrt{3}$, that is half of the 
diagonal length of the whole system under study. 
Taking into account that we are using a minimum image convention to measure
distance, this is the maximum available distance in our simulated cell.
Hence, we conclude that the saturation takes place when the defect lines
cannot continue to grow. The thermal creation of new defects 
cannot increase the length of the cluster due to the finite size effects.

Let us now demonstrate that the chains of defects studied in this work must 
in fact be dislocation lines.
Consider a plane which is orthogonal to the chain of defects and contains one 
of the vacancy-interstitial pairs that the chains is formed of. Any symmetric 
contour around this vacancy-interstitial pair in that plane does not close, 
a very well-known fact in 2D; its misfit is the Burgers vector of 
a dislocation formed by this pair of defects. Since the Burgers vector 
magnitude of a dislocation in 2D is related to the separation of 
the underlying defects, the contour misfits in our 3D case will be the same 
for every plane orthogonal to the array and containing a vacancy-interstitial 
pair: the lines of vacancies and interstitials are parallel to each other, 
hence the defect separation is the same for every plane. Thus, the orthogonal 
contour misfit is an invariant for every chain of defects: it is in fact 
the (nonzero) Burgers vector of a dislocation formed by this array.

What is the seed of the melting process? Two different scenarios based on
the dislocation-line generation could be envisaged. The first one, invoked 
by most of the existing theories, consider that melting takes place when the 
crystal is saturated with dislocation loops of  
all sizes including open dislocation lines crossing the whole system. 
In fact, the probability to have a loop of length $\ell $ in 
a crystal at the critical point is $p(\ell )\sim \ell ^{-q},$ where 
the exponent $q$ depends on the line topology, in other words, the nature of 
the dislocations as random walks, i.e., Brownian, self-avoiding, etc., and 
the balance between the loops and open lines \cite{Bura}. The other 
possibility, see Ref.\ \onlinecite{Vesh}, is to associate melting with 
the generation of an arbitrary low density of infinitely long dislocation 
dipoles. These dipoles are the pairs of dislocations with opposite Burgers 
vectors. As seen in Fig.\ \ref{CNvsT}, the density of the defective atoms is 
rather high, $\sim 40$\% at the critical point, which is inconsistent with 
the density of dislocation dipoles being (arbitrary) low, according to 
the second scenario. (In fact, this percentage of the defective atoms is 
consistent with the conclusion of Ref.\ \cite{Bura} that about 1/2 of all 
atoms are within the dislocation cores at the critical point.) It is 
therefore plausible to suggest that the first scenario is realized in our 
simulation, i.e., the proliferation of dislocation loops of all sizes, 
including the appearance of very long open dislocation lines near $T_m$. 

To gain further support for this scenario, we have analyzed in detail 
how the clusters of defects are formed at temperatures very near melting.

In Fig.\ \ref{NvsL}c and \ref{NvsL}d 
we show a plot of $L$ vs.\ $N$ at $T_1=0.52$ and $T_2=0.57$, 
i.e., both in the premelting zone but $T_2$ near $T_m$. The most prominent 
difference between these two figures is that at $T_1$ the bigger cluster 
could include any number of particles beyond a critical one. However, at 
$T_2$ the bigger clusters appear with a specific number of defects inside it.
This seems to indicate that these clusters are the ensemble of lines that 
cross the whole system. In addition, there is one of these big clusters in 
each realization of our simulation, and {\it all} the defective particles 
belong to this cluster at that temperature \cite{NdefNline}. According to 
our results, such big clusters of defects are the precursor of melting. 
In fact, we have found that some of the samples that contain these big 
clusters melt after a long simulation. 

The mechanism of 
the clustering of point defects into dislocation loops in the premelting 
stage seems to have certain experimental support. Novikov {\it et al.} 
\cite{Novikov} suggested that with their concentration growing, thermally 
generated point defects must cluster into dislocation loops for energetic 
reasons, and that this effect will be observed in a crystal in the state of 
premelting. They themselves claim the observation of this effect in lead. 
Two very recent experiments lend additional support: in Ref.\ \cite{Gondi} 
the point defect clustering into dislocation loops was observed in indium 
during its melting, and in Ref.\ \cite{HNO} the formation of defect 
clusters was observed in silicon close to its melting temperature. 

To summarize, we have analyzed the thermal excitation of defects by Monte 
Carlo simulation. Beyond a crossover temperature, defects group in clusters of 
all sizes that correspond to dislocation loops. Near the melting temperature, 
we observe the appearance of very long lines of defects that cross the whole 
system. We identify these lines with the precursor of the melting transition. 

I Electronic address: liliana@fceia.unr.edu.ar\\
II Electronic address: dobry@fceia.unr.edu.ar\\
III Electronic address: diep@ptm.u-cergy.fr\\
IV Electronic address: burakov@lanl.gov \\

\end{document}